\documentclass[aps,pra,reprint,nofootinbib,showpacs,superscriptaddress]{revtex4-1}
\input{Qcircuit}
\usepackage{graphicx} 
\usepackage{enumerate}
\usepackage{hyperref}
\usepackage{amsfonts}
\usepackage{amsmath,amssymb}
\usepackage{mathrsfs}
\usepackage{amsthm}
\usepackage{dsfont}
\usepackage{bm} 
\usepackage{color}
\usepackage{epstopdf} 
\usepackage{epsfig}

{\rm }

\def\be{\begin{equation}}
 \def\ee{\end{equation}}
 \def\bea{\begin{eqnarray}}
 \def\eea{\end{eqnarray}}
 \def\bes{\begin{eqnarray}}
 \def\ees{\end{eqnarray}}
 \def\bi{\begin{itemize}}
 \def\ei{\end{itemize}} 

\renewcommand{\sec}[1]{\hyperref[sec:#1]{Sec.~\ref{sec:#1}}}

\newcommand{\fig}[1]{\hyperref[fig:#1]{Fig.~\ref{fig:#1}}}

\def\H{\mathcal{H}}

\def\2{\frac{1}{2}}
\def\4{\frac{1}{4}}

\begin{document}

\title{Physical realization of topological quantum walks on IBM-Q and beyond}
+

\author{Radhakrishnan Balu}
\email{radhakrishnan.balu.civ@mail.mil}
\affiliation{Computer and Information Sciences
	Directorate, Army Research Laboratory, Adelphi, MD 21005-5069.}
\affiliation{Computer Science and Electrical Engineering,
              University of Maryland Baltimore County,
              1000 Hilltop Circle, Baltimore, MD 21250.}
\email{radbalu1@umbc.edu}
\author{Daniel Castillo}
\email{dcastil1@vols.utk.edu}\affiliation{Department of Physics and Astronomy, The University of Tennessee, Knoxville, TN 37996-1200, U.S.A.}
\author{George Siopsis}
\email{siopsis@tennessee.edu}
\affiliation{Department of Physics and Astronomy, The University of Tennessee, Knoxville, TN 37996-1200, U.S.A.}

\date{\today}

\begin{abstract}
	
	We discuss an efficient physical realization of topological quantum walks on a finite lattice. The $N$-point lattice is realized with $\log_2 N$ qubits, and the quantum circuit utilizes a number of quantum gates which is polynomial in the number of qubits. In a certain scaling limit, we show that a large number of steps is implemented with a number of quantum gates which is independent of the number of steps. We ran the quantum algorithm on the IBM-Q five-qubit quantum computer, thus experimentally demonstrating topological features, such as boundary bound states, on a lattice with $N=4$ points. 
	
\end{abstract}

\maketitle
\onecolumngrid
\section{Introduction}
Quantum walks form a versatile framework for studying myriad of physical processes ranging from biological systems \cite{Mohseni2008}, to satisfiability problems in computer science \cite {FAHRI2012}, to topological systems in physics \cite{Kitagawa2012,Hasan2010,Qi2011} as they are shown to perform universal computation \cite{Child2013}. The long-time limit of usual quantum walks results in an arc-sine law, and in one case Konna \cite{Konna2014} has constructed quantum walks whose asymptotic limits are uniform measures. Recent studies have established a relationship between the Dirac equation and discrete-time quantum walks by observing the causal evolution as the basis for relativistic features \cite{Chand2012,Strauch,Strauch2006}. For an introduction we refer the reader to the work of Salvador \cite{Salvador2012} that has a comprehensive review on the quantum walk formalism and its many possible applications, and \cite{Kempe2003} for a gentle
introduction to this topic. Quantum walks interacting with an environment were studied by Attal, Petruccione, and Sinayskiy \cite{Attal2012,Petruccione2013} as an open quantum system. Continuous-time quantum walks as a limit of their discrete-time versions were formulated in the one-dimensional case by Debbasch, \emph{et al.}\ \cite{Deb2012}. Liu, \emph{et al.}\ described continuous walks  on a graph in open quantum evolutions \cite{Rad2017}. Quantum walks can be formulated with a Hamiltonian exhibiting chiral symmetry that are similar to open-system evolutions in terms of time-symmetry breaking features \cite{Lu2016}. Cardona, \emph{et al.}, \cite{Cardona2015} realized experimentally quantum walks on orbital angular momentum space of light, and Obuse, \emph{et al.}, have realized walks based on Hadamard coins \cite{Obuse2015}. Other proof-of-principle demonstrations of quantum walks have been experimentally realized using single photons \cite{Schreiber2010,Broome2010,Peruzzo2010,Schreiber2012,Sansoni2012,Jeong2013}, trapped ions \cite{Zahringer2010,Schmitz2009}, and atoms in optical lattices \cite{Karski2009,Genske2013}.

Topological quantum walks have been extensively studied both theoretically\cite{Obuse2011,Asboth2012,Asboth2013,Ramasesh2016,Rakovszky2015}, and experimentally \cite{Cardano2016}. One way to create non-trivial topological phases is to fashion the quantum walks in split steps using two coins with a requirement that the rotation of the second coin is inhomogeneous \cite{Kitagawa2012b}. This prescription has been realized experimentally using photons on an optical lattice \cite{Kitagawa2012} where they performed a stroboscopic simulation of an effective Hamiltonian.  In this work, we take the continuous-time limit of discrete topological quantum walks \cite{Strauch,Strauch2006} with split-step evolutions, identify their phases and boundaries, and realize them using quantum circuits. An alternate continuous-limit approach given in \cite{Dheeraj2015} is not considered here, although it may be interesting for future studies of topological properties of quantum walks. Whereas stroboscopic simulations are useful in realizing effective Hamiltonians, highlighting the effectiveness of quantum walks as simulation tools, they require a lot of resources in terms of current technology. Our approach uses the limited quantum hardware available to establish the connection between quantum walks and their continuous-time-limit  dynamics. One way to understand the localized states is to look at the effective Hamiltonian and its dispersion that provide band gaps for certain parameters of the rotations. Accordingly, we chose the parameters for which there is a definite gap so that a walk starting in one phase stays at the same phase. When the initial state is at the boundary of two topologically distinct phases, it remains there (bound state). We implemented the simplest topological quantum walks on the IBM-QX4 quantum computer consisting of five qubits.

Our discussion is organized as follows. In section \ref{sec:2}, we introduce topological quantum walks on a one-dimensional lattice, and discuss their efficient implementation with quantum circuits. In section \ref{sec:3}, we discuss a scaling (continuous-time) limit that allows the implementation with quantum circuits whose complexity does not grow with the number of steps of the walker. We discuss the quantum circuits and simulate them. In section \ref{sec:4}, we present experimental results from the physical realization of the simplest topological quantum walks on IBM-Q, demonstrating the existence of bound states. Finally, in section \ref{sec:5}, we summarize our conclusions.

\section{Topological quantum walk}
\label{sec:2}

For definiteness, we concentrate on split-step topological quantum walks \cite{Kit} on a lattice of $N = 2^n$ points forming a one-dimensional circle. The discussion can easily be extended to more general lattices. Thus, the quantum walk consists of a walker and coin Hilbert spaces, $\H_w=\{\ket{x}, x\in \mathbb{Z}_N \}$, and $\H_c=\{ \ket{0}, \ket{1}\}$, respectively. Applying periodic boundary conditions, we define the shift operators as
\be\label{eqL0}
L^\pm \ket{x}_w = \ket{(x\pm 1)\mod N}_w
\ee
in terms of which the step operators are
\be\label{S}
S^+ = \ket{0}_c\bra{0}\otimes L^+ + \ket{1}_c\bra{1}\otimes \mathbb{I}\ , \ \
S^- = \ket{0}_c\bra{0}\otimes \mathbb{I} + \ket{1}_c\bra{1}\otimes L^-
\ee
The coin operator is
\be\label{eqT}
T(\theta) = e^{-i\theta Y} =
\left(\begin{matrix}
	\cos\theta & -\sin\theta\\
	\sin\theta & \cos\theta
\end{matrix}\right)
\ee
where $\theta$ varies with the position $x$ on the lattice.
A single step of the quantum walk is given by the unitary
\be\label{eqW}
W = S^-T(\theta_2)S^+T(\theta_1)
\ee
in terms of a pair $(\theta_1, \theta_2)$ of coin parameters. This system possesses topologically distinct phases for different values of the coin parameters \cite{Kit}. We will discuss the case in which the lattice consists of two regions in two different topological phases forming a boundary between them on which bound states may form. The coin parameters will have different values in the two regions denoted by
\be\label{eq5}
(\theta_1, \theta_2) = 
\left\{ \begin{array}{ccc}
	(\theta_{1}^-, \theta_{2}^{-}) &, & 0\le x<\frac{N}{2}\\
	(\theta_{1}^{+}, \theta_{2}^{+}) &, & \frac{N}{2} \le x < N
\end{array}\right.
\ee
Thus, the two coin operators are of the form
\be \mathcal{T}_i = \sum_{x=0}^{N-1} e^{-i\theta_i Y_c} \otimes |x\rangle_w\langle x| \ , \ \ i=1,2\ee
The single step of the quantum walk \eqref{eqW} becomes
\be\label{eqWa}
W = S^-\mathcal{T}_2S^+\mathcal{T}_1
\ee
We will use $n+1$ qubits to realize the system, $n$ for $\mathcal{H}_w$ and $1$ for $\mathcal{H}_c$. To implement $L^\pm$, notice that the states
\be\label{eqk} |\overline{k}\rangle_w = \frac{1}{\sqrt{N}} \sum_{x=0}^{N-1} e^{-2\pi ikx/N} |x\rangle_w  \ee
are eigenstates of the shift operators,
\be L^\pm |\overline{k}\rangle_w = e^{\pm 2\pi ik/N} |\overline{k}\rangle_w \ee
Therefore,
\be\label{eqL} L^\pm = \sum_{k=0}^{N-1} e^{\pm 2\pi ik/N} |\overline{k}\rangle_w \langle \overline{k}| \ee
The states $|\overline{k}\rangle_w$ and $|x\rangle_w$ are related to each other via the Quantum Fourier Transform (QFT),
\be |\overline{k}\rangle_w = U_{QFT} |k\rangle_w \ , \ \ \langle x|U_{QFT} |k\rangle =  \frac{1}{\sqrt{N}} e^{-2\pi ikx/N} \ee
Eq.\ \eqref{eqL} can be written in terms of the QFT as
\be\label{L2} L^\pm =  U_{QFT}\prod_{j=0}^{n-1} e^{\mp \pi i Z_j/2^{j+1}}  U_{QFT}^\dagger \ee
Consequently, the step operators, $S^\pm$, given by \eqref{S}, become
\be S^\pm = U_{QFT} \mathcal{S}^\pm   U_{QFT}^\dagger \ , \ \
\mathcal{S}^\pm = \prod_{j=0}^{n-1} e^{\pi i(Z_c\mp Z_j) /2^{j+2}} e^{-\pi i Z_cZ_j/2^{j+2}}
\ee
where the two-qubit gates can be written in terms of CNOT gates as
\be e^{-i\theta Z_cZ_j} = \text{CNOT}_{jc} \cdot e^{-i\theta Z_c} \cdot \text{CNOT}_{jc} \ee
An example of a quantum circuit implementing $\mathcal{S}^\pm$ for $n=2$ is shown below:
\[
\mathcal{S}^\pm: \ \ \ \ \ 
\Qcircuit @C=1em @R=.7em @!R {
	& \lstick{c\ \ } & \lstick{\cdots} & \targ & \gate{e^{-i\pi Z/8}} & \targ & \targ & \gate{e^{-i\pi Z/4}} & \targ & \gate{e^{3i\pi Z/8}} & \qw &\cdots \\
	& \lstick{0\ \ } & \lstick{\cdots} & \ctrl{-1} & \qw & \ctrl{-1} & \qw & \qw & \qw & \gate{e^{\mp i\pi Z/8}} & \qw &\cdots \\
	& \lstick{1\ \ } & \lstick{\cdots} & \qw & \qw & \qw & \ctrl{-2} & \qw & \ctrl{-2} & \gate{e^{\mp i\pi Z/4}} & \qw &\cdots
}
\]
Since $\theta_i$ ($i=1,2$) depends only on the first digit in the binary expansion of $x$, $\mathcal{T}_i$ is implemented with a single qubit operation on the coin and CNOT with control the $j=0$ qubit and target the coin,
\be \mathcal{T}_i =  e^{-i\frac{\theta_i^++\theta_i^-}{2} Y_c} \cdot \text{CNOT}_{0c} \cdot e^{-i\frac{\theta_i^+-\theta_i^-}{2} Y_c}  \cdot \text{CNOT}_{0c}\ee
The quantum circuit for $\mathcal{T}_i$ is shown below in the dotted box:
\[
\mathcal{T}_i: \ \ \ \ 
\Qcircuit @C=1em @R=.7em @!R {
	& \lstick{c\ \ } & \lstick{\cdots} & \gate{e^{-i\frac{\theta_{i}^{+}+\theta_{i}^{-}}{2}Y}} & \targ & \gate{e^{-i\frac{\theta_{i}^{+}-\theta_{i}^{-}}{2}Y}} & \targ & \qw & \rstick{\cdots} \\
	& \lstick{0\ \ } & \lstick{\cdots} & \qw & \ctrl{-1} & \qw & \ctrl{-1}  & \qw & \rstick{\cdots} \\
	& \lstick{\ \ } & \lstick{\cdots} & \qw & \qw & \qw & \qw &  \qw & \rstick{\cdots} \gategroup{1}{4}{2}{7}{.7em}{--}
}
\]
The topological effects are captured by the CNOT gates; the latter are absent in the case of a single topological phase.

For the repeated step of the walker \eqref{eqWa}, the quantum circuit for $n=2$ is shown below:
\be\label{eqcW}
W: \ \ \ \ 
\Qcircuit @C=1em @R=.7em @!R {
	& \lstick{c\ \ } & \lstick{\cdots} & \multigate{1}{\mathcal{T}_{1}}  &  \qw & \multigate{2}{\mathcal{S}^+} & \qw & \qw& \multigate{1}{\mathcal{T}_{2}}  &  \qw & \multigate{2}{\mathcal{S}^-} & \qw & \qw & \rstick{\cdots} \\
	& \lstick{0\ \ } & \lstick{\cdots} & \ghost{\mathcal{T}_{1}} &  \multigate{1}{U^\dagger_{QFT}} & \ghost{\mathcal{S}^+} & \multigate{1}{U_{QFT}} & \qw & \ghost{\mathcal{T}_{2}} &  \multigate{1}{U^\dagger_{QFT}} & \ghost{\mathcal{S}^-} & \multigate{1}{U_{QFT}} & \qw & \rstick{\cdots} \\
	& \lstick{1\ \ } & \lstick{\cdots} & \qw  & \ghost{U_{QFT}} & \ghost{\mathcal{S}^+} & \ghost{U^\dagger_{QFT}} & \qw& \qw  & \ghost{U_{QFT}} & \ghost{\mathcal{S}^-} & \ghost{U^\dagger_{QFT}} & \qw & \rstick{\cdots} 
} \ee

\section{Scaling limit}
\label{sec:3}
The quantum circuits considered above require a number of quantum gates which is proportional to the number of steps of the walker. Thus, they require a considerable amount of quantum resources, if we want to understand the topological quantum walk at a large number of steps. To study a large number of steps, we consider a scaling limit that leads to quantum circuits that do not grow in length as we increase the number of steps. In this limit, the coin parameters $(\theta_1,\theta_2)$ approach the following values in the four distinct topological phases, respectively,
\be\label{ph2}
\text{I:}\ \ \left( 0, \frac{\pi}{2} \right) \ , \ \
\text{II:}\ \   \left( 0, -\frac{\pi}{2} \right) \ , \ \
\text{III:}\ \  \left( \frac{\pi}{2},0 \right) \ , \ \
\text{IV:}\ \   \left( -\frac{\pi}{2},0 \right)
\ee
Consider phase I first. 
We set $\theta_1 = \epsilon$ and $\theta_2 = \frac{\pi}{2}+\epsilon$, and take the limit as $\epsilon\rightarrow 0$ and the number of steps $s\rightarrow\infty$ while the product $s\epsilon$ remains finite. Since $W^4 = \mathbb{I} + \mathcal{O}(\epsilon)$, we consider a number of steps which is multiple of 4, and set
\be\label{eq17} s \epsilon = \frac{t}{2} \ee
thus trading the discrete variable $s$ for the continuous (time) variable $t$. For the state of the system $|\Psi(t)\rangle$,
we obtain equations of motion of the Schr\"odinger form for phase I,
\be\label{eqI} i\frac{d}{dt} |\Psi\rangle = H_{\text{I}} |\Psi\rangle \ , \ \
H_{\text{I}} = - Y \otimes \left[ 2\mathbb{I} + L^+ +L^- \right]
\ee
where $Y$ is a Pauli matrix acting on the coin.

Working similarly, we obtain equations of motion of the Schr\"odinger form for the other phases with Hamiltonians, respectively,
\bea H_{\text{II}} &=& -Y\otimes \left[ 2\mathbb{I} - L^+ - L^-\right]
\ , \nonumber\\ H_{\text{III}} &=& -Y\otimes \mathbb{I} +i X^+ \left[ L^+ + (L^+)^2 \right] - iX^-\otimes \left[ L^- + (L^-)^2 \right]
\ , \nonumber\\ H_{\text{IV}} &=& Y\otimes \mathbb{I} +i X^+ \left[ L^+ - (L^+)^2 \right] - iX^-\otimes \left[ L^- - (L^-)^2 \right]
\eea
where $X^\pm = X \pm i  Y$.
We deduce the evolution of the system in a single topological phase,
\be |\Psi(t)\rangle = e^{-iHt} |\Psi(0)\rangle \ee
where $H$ is a $2n\times 2n$ sparse matrix that can be efficiently simulated. 


As an example, consider phase I in the case $N=4$. The Hamiltonian is a $8\times 8$ Hermitian matrix that
can be written in terms of Pauli matrices as
\be
H_{\text{I}} = Y\otimes \left( 2\mathbb{I}\otimes\mathbb{I}+\mathbb{I}\otimes X +  X\otimes X \right)
\ee
Since all terms in the expansion commute with each other, we can factor the evolution operator as
\be\label{singlephasetd}
e^{-iH_{\text{I}} t} =  e^{-2i t Y_c}e^{-i t Y_cX_1}e^{ - i t Y_cX_0X_1}
\ee
and implement it with CNOT and single qubit gates. The quantum circuit for \eqref{singlephasetd} is
\be\label{cSinglephasetd}
\Qcircuit @C=1em @R=.7em @!R {
	& \lstick{c\ \ } & \ctrl{2} & \gate{e^{-i t Y/2}} & \ctrl{1} & \gate{e^{-it Y/2}} & \ctrl{1} & \ctrl{2} & \gate{e^{-i t Y}} & \qw \\
	& \lstick{0\ \ } & \qw & \qw & \targ & \qw & \targ & \qw  & \qw & \qw \\
	& \lstick{1\ \ } & \targ & \qw & \qw & \qw & \qw & \targ  & \qw & \qw
}
\ee
\subsubsection{Example: d = 8}

\begin{figure}[htp]
	\centering
	\includegraphics{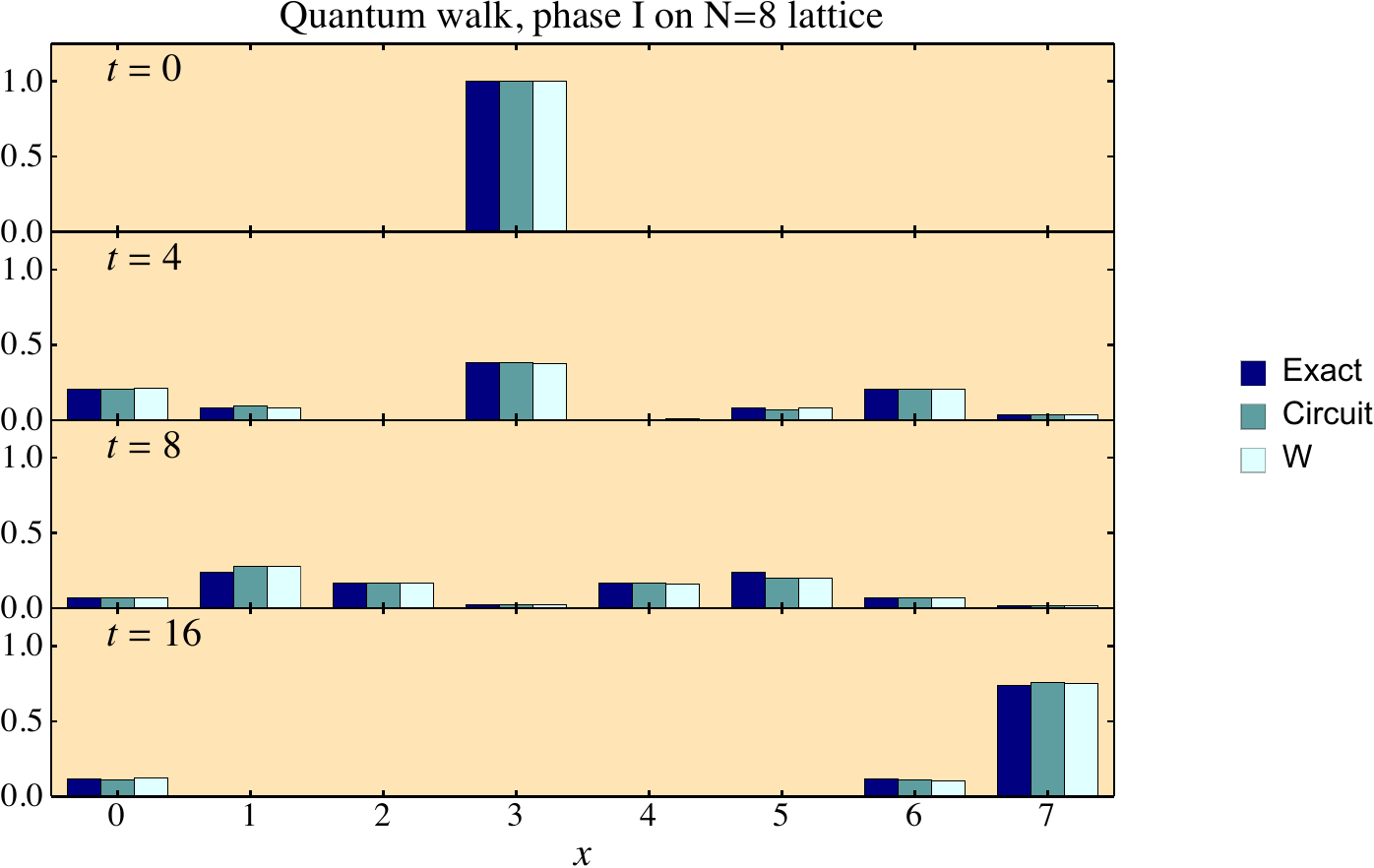}
	\caption{Quantum walk in topological phase I on a lattice of $N=8$ points with coin parameters $\theta_1=\epsilon,\theta_2=\frac{\pi}{2}+\epsilon$, where $\epsilon=\frac{1}{8}$. Numerical results using the exact Hamiltonian \eqref{eq24} are compared with results from the quantum circuit for \eqref{N8singlephasetd} at various values of $t$, and repeated application of the matrix \eqref{eqW} representing a single step of the walker for number of steps $s$ related to $t$ via \eqref{eq17}.}\label{fig:N8NonTop}
\end{figure}

For $N=8$, we obtain similarly,
\be\label{eq24}
H_{\text{I}} = Y \otimes \left( 2\mathbb{I} \otimes \mathbb{I} \otimes \mathbb{I} +\mathbb{I}\otimes \mathbb{I} \otimes X + \frac{1}{2} \mathbb{I} \otimes X \otimes X + \frac{1}{2} \mathbb{I}\otimes Y\otimes Y+\frac{1}{2} X \otimes X\otimes X - \frac{1}{2} X\otimes Y\otimes Y \right)
\ee
For small $t$, the evolution operator is
\be\label{N8singlephasetd}
e^{-iH_{\textbf{I}} t} =  e^{-2i t Y_c}e^{-i t Y_cX_2}e^{-\frac{i}{2} t Y_cX_1X_2}e^{-\frac{i}{2} t Y_cY_1Y_2}e^{-\frac{i}{2} t Y_cX_0X_1X_2}e^{\frac{i}{2} t Y_cX_0Y_1Y_2} + \mathcal{O} (t^2)
\ee
This expression is a good approximation in the case of a large number of walker steps ($s\gg 1$), if we choose a small enough coin parameter $\epsilon$ ($s\epsilon \ll 1$). For a larger value of $\epsilon$, we need to split the time in the evolution operator,
\be e^{-iH_{\text{I}}t} = \lim_{T\to\infty} \left( e^{-iH_{\text{I}}t/T} \right)^T \ee
and apply \eqref{N8singlephasetd} on each factor $e^{-iH_{\text{I}}t/T}$.
Eq.\ \eqref{N8singlephasetd} can be implemented as a quantum circuit in terms of CNOT and single qubit gates. Numerical results from a simulation of the circuit for the evolution \eqref{N8singlephasetd} for various values of the time parameter $t$ are shown on figure \ref{fig:N8NonTop}. The initial state is chosen with support at the point $x=3$. The coin parameters were set to $\theta_1=\epsilon,\theta_2=\frac{\pi}{2}+\epsilon$, with $\epsilon=\frac{1}{8}$. According to eq.\ \eqref{eq17}, $t$ corresponds to number of steps $s= 4t$. The results from the quantum circuit are compared with those from the evolution $e^{-itH_\text{I}}$ using the exact expression \eqref{eq24} for the Hamiltonian, and found in good agreement with each other. They are also in agreement with results from a repeated application of the unitary $W$ representing a single step of the walker, as expected for small $\epsilon$.


The above matrices are modified when two or more distinct topological phases are present because of boundary effects. Let us consider the case in which the system is in phase II (I) for $0\le x < \frac{N}{2}$ ($\frac{N}{2} \le x < N$). In the notation of eq.\ \eqref{eq5}, we choose $\theta_1^\pm = \epsilon$, $\theta_2^+ = - \theta_2^- = \frac{\pi}{2} + \epsilon$. Working as before, for $N=4$ we obtain the Hamiltonian
\be\label{se}
 H_{\text{I/II}} 
= Y \otimes \mathbb{I}\otimes \frac{\mathbb{I} +Z}{2}
\ee
Due to the projection matrix $\frac{\mathbb{I}+Z}{2}$, it has a four-dimensional kernel, therefore four bound states at points $x=1,3$.
The evolution matrix $e^{-it H_{\text{I/II}}}$ can be implemented with the quantum circuit
\be\label{eqcC}
\Qcircuit @C=1em @R=.7em @!R {
	& \lstick{c\ \ } & \gate{e^{-itY/2}} & \targ & \gate{e^{-itY/2}} & \targ & \qw \\
	& \lstick{0\ \ } & \qw & \qw & \qw & \qw & \qw \\
	& \lstick{1\ \ } & \qw &  \ctrl{-2} & \qw & \ctrl{-2} & \qw
}
\ee

\begin{figure}[htp]
	\centering
	\includegraphics{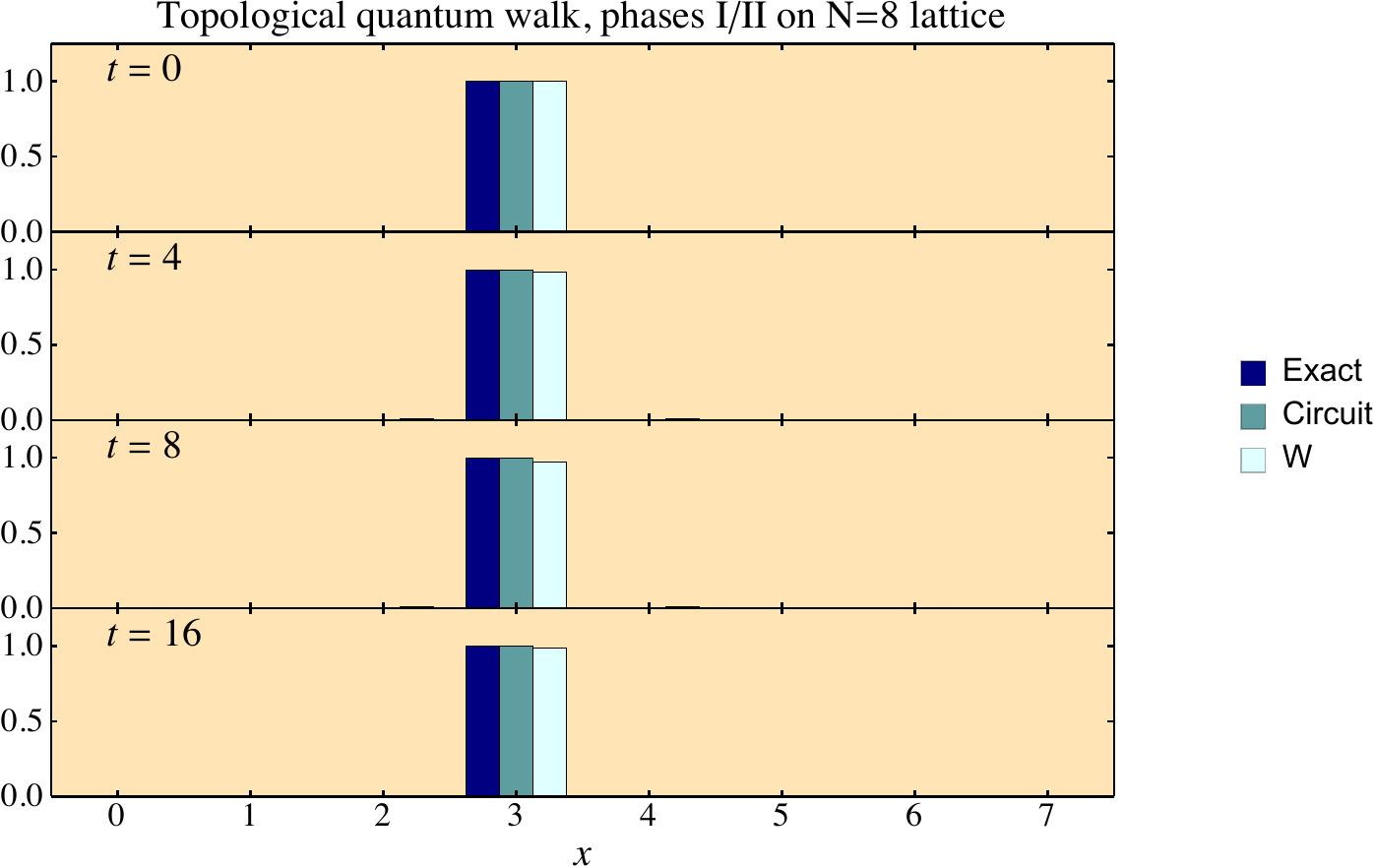}
	\caption{Topological quantum walk in phases I/II on a lattice of $N=8$ points with coin parameter $\epsilon=\frac{1}{8}$. Initial state at the boundary point $x=3$.
Numerical results using the exact Hamiltonian \eqref{eq29} are compared with results from the quantum circuit for \eqref{N8singlephasetop} at various values of $t$, and repeated application of the matrix \eqref{eqW} representing a single step of the walker for number of steps $s$ related to $t$ via \eqref{eq17}.
		}\label{fig:N8TimeDep}
\end{figure}

Similarly, for $N=8$ we obtain the two-phase Hamiltonian
\be\label{eq29}
H_{\text{I/II}} = \frac{1}{2} Y_c \left( 3 \mathbb{I} + Z_1 + Z_2 - Z_1Z_2 + Z_0X_1X_2 + Z_0Y_1Y_2 + Z_0X_2 + Z_0Z_1X_2 \right)
\ee
It has bound states at the boundary points $x=3,7$, as is easily verified.
As in the single-phase case, the evolution operator can be approximated for small $t$ by
\be\label{N8singlephasetop}
e^{-itH_{\text{I/II}} } =  e^{-3i t Y_c/2}e^{i t Y_cZ_1Z_2/2}e^{-i t Y_cZ_2/2}e^{-i t Y_cZ_1/2} e^{-i t Y_cZ_0X_1X_2/2}e^{-i t Y_cZ_0Y_1Y_2/2}e^{-i t Y_cZ_0X_2/2}e^{-i t Y_cZ_0Z_1X_2/2} + \mathcal{O} (t^2)
\ee
and implemented with a quantum circuit in terms of CNOT and single qubit gates. In figure \ref{fig:N8TimeDep}, we compare numerical results from a simulation of the quantum circuit for \ref{N8singlephasetop} with the evolution using the exact Hamiltonian \eqref{eq29}, as well as a repeated application of a single walker step $W$, with the number of steps related to the time parameter $t$ via \eqref{eq17}. For the choice of coin parameter $\epsilon = \frac{1}{8}$, we obtain good agreement between the different approaches. 
The results demonstrate the existence of a bound state at the boundary point $x=3$ due to topological effects. This should be contrasted with the results depicted in figure \ref{fig:N8NonTop} for a quantum walk in phase I, where the original peak at $x=3$ dissipates after a finite number of steps of the walker.

The above results can be easily extended to other combinations of topological phases, as well as larger walker Hilbert spaces $\mathcal{H}_w$. 
In general, the system possesses localized bound states at the boundary points $x\approx \frac{N}{2}$ and $x\approx N$. For large enough $N$, when the walker is sufficiently far from the boundaries, the equations of motion reduce to those for a single phase.

\section{IBM-Q}
\label{sec:4}
\begin{figure}[htp]
	\centering
	\includegraphics[width=\textwidth]{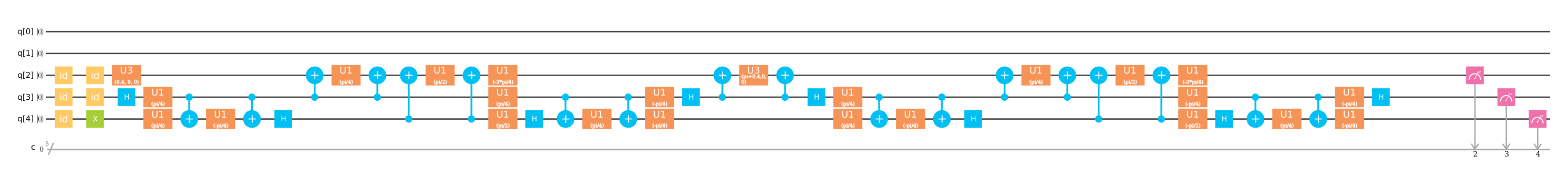}
	\caption{The quantum circuit for a single step $W$ (eq.\ \eqref{eqcW}) of a topological quantum walk in phases I/II \eqref{cSinglephasetd} for $N=4$ implemented on IBM-Q. For phase I, remove the CNOT gates before and after the coin-flip gates. In the circuit, $\textsf{U3(u,0,0)} = e^{-iu Y}$, $\textsf{U1(u)} =e^{-iu Z}$.}\label{fig:OneStepIBM}
\end{figure}

\begin{figure}[ht]
	\centering
	\includegraphics{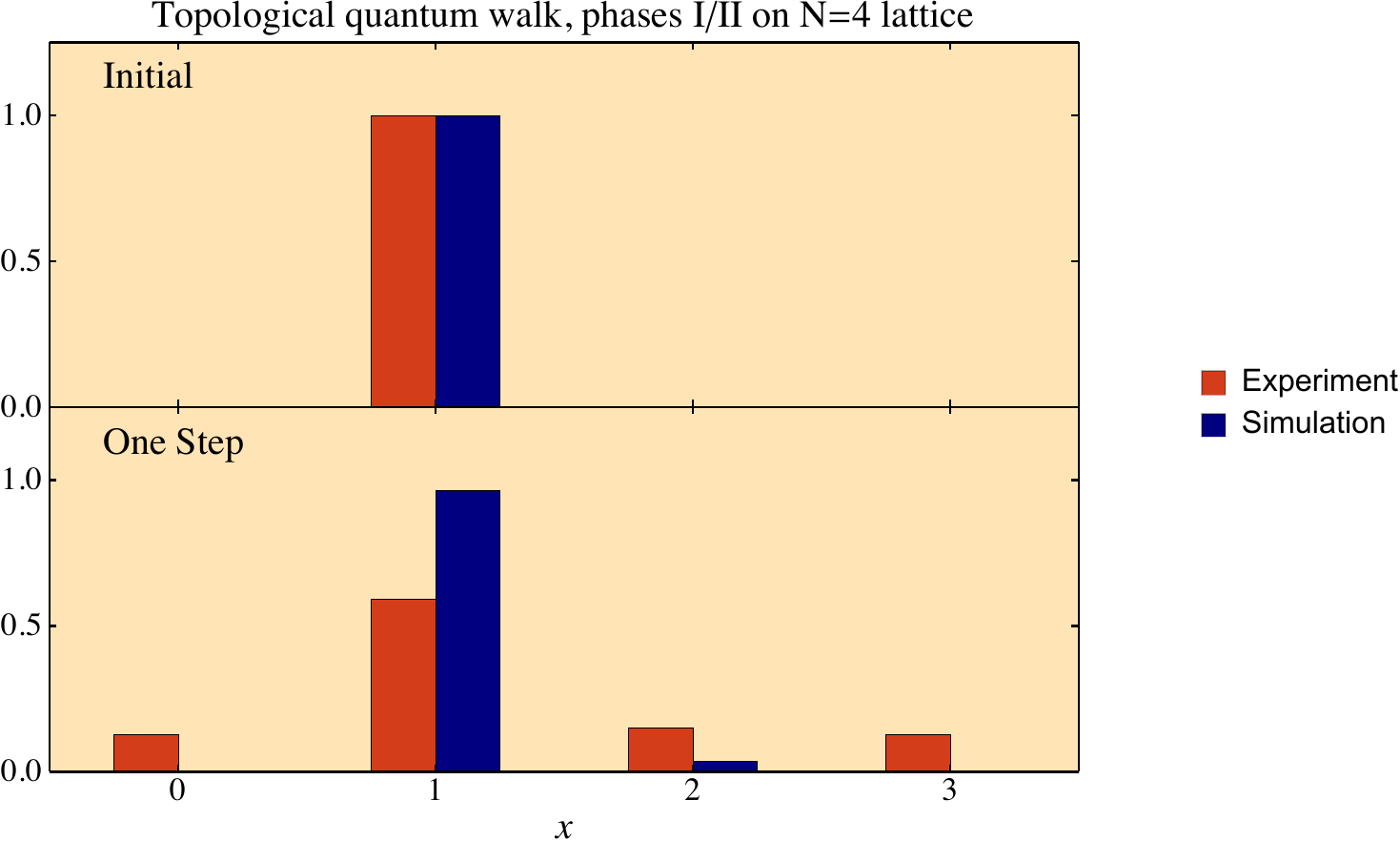}
	\caption{Results for a single step $W$ (eq.\ \eqref{eqcW}) of a topological quantum walk in phases I/II \eqref{cSinglephasetd} for $N=4$ implemented on IBM-Q. The initial state is chosen so that it remains bound. Coin parameter set at $\epsilon = \frac{1}{8}$. Experimental results are compared with simulation.}\label{fig:OneStep}
\end{figure}

\begin{figure}[ht]
	\centering
	\includegraphics[width=\textwidth]{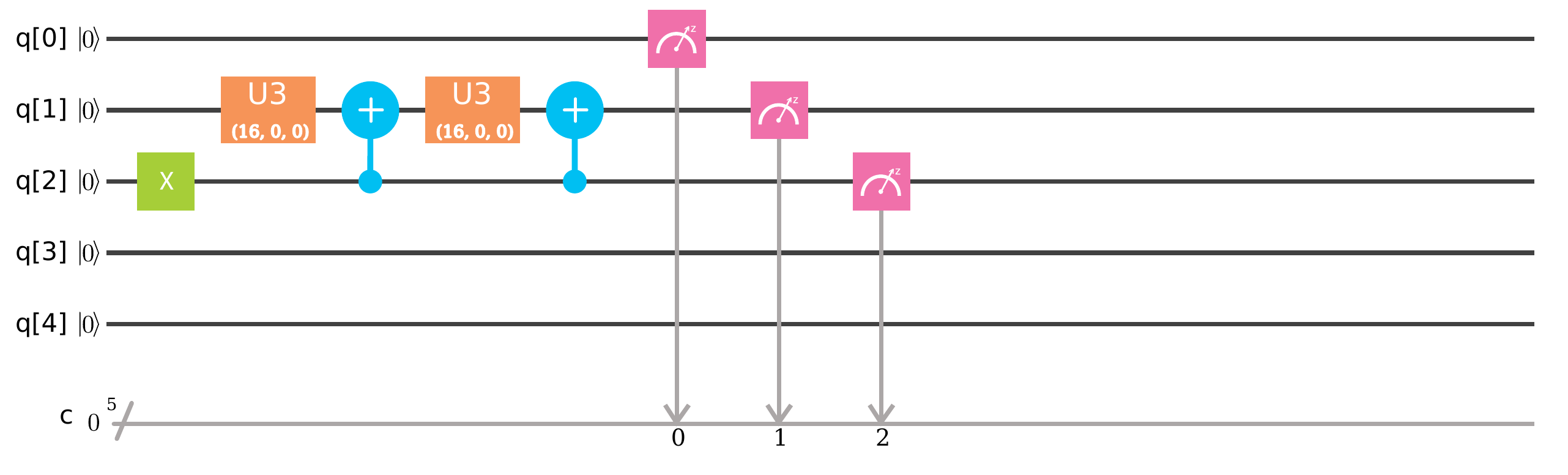}
	\caption{The quantum circuit for a large number of steps in the scaling limit (eq.\ \eqref{eqcC}) of a topological quantum walk in phases I/II for $N=4$ implemented on IBM-Q. In the circuit, $\textsf{U3(t,0,0)} = e^{-it Y}$.}\label{fig:ContIBM}
\end{figure}

\begin{figure}[ht]
	\centering
	\includegraphics{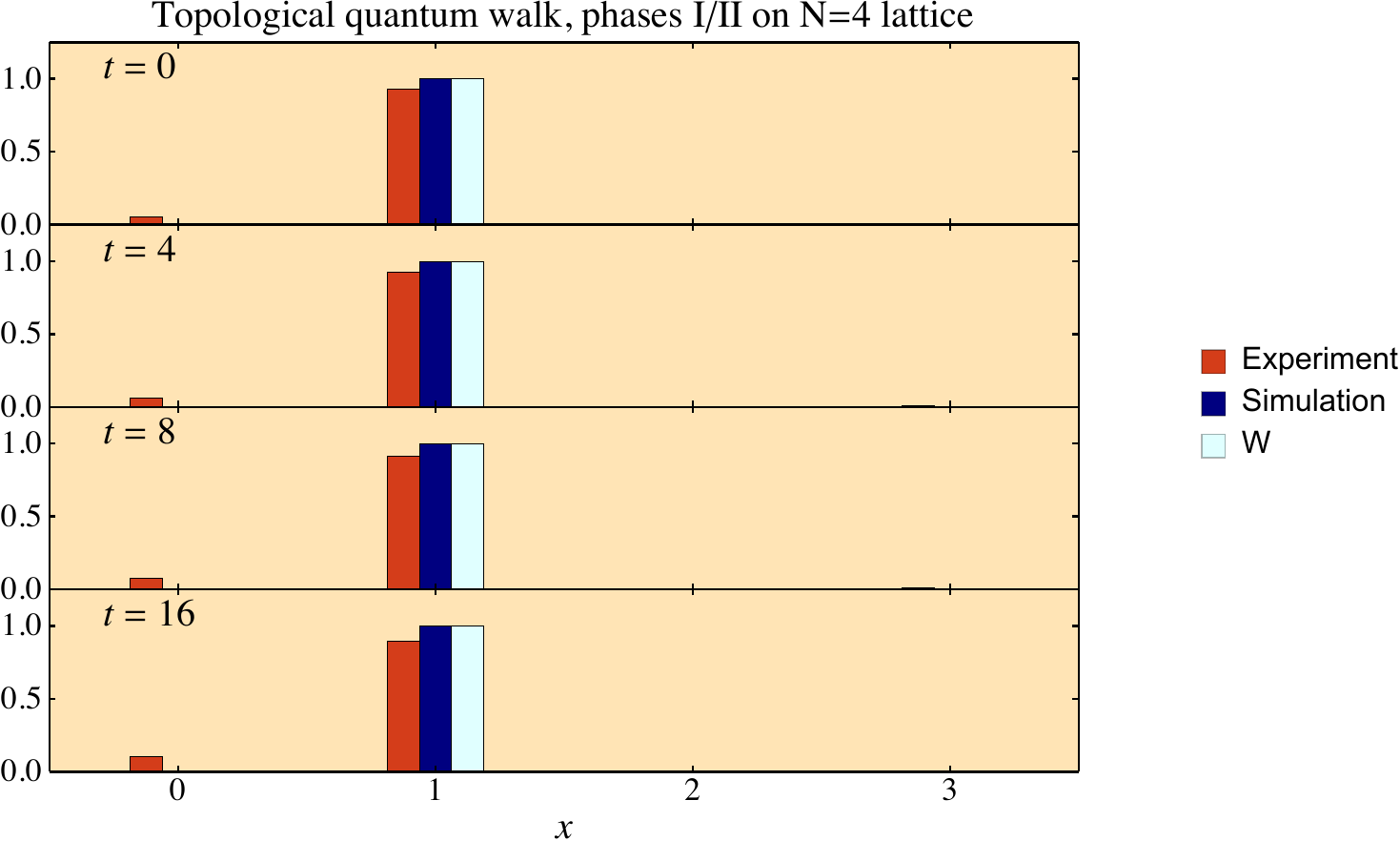}
	\caption{Topological quantum walk in phases I/II on a lattice of $N=4$ points with coin parameter $\epsilon = \frac{1}{32}$. Experimental results from IBM-Q demonstrate the existence of a bound state at $x=1$. They compare well with simulation, and repeated application of the matrix \eqref{eqW} representing a single step of the walker for number of steps $s$ related to $t$ via \eqref{eq17}.}\label{fig:TimeDep}
\end{figure}
\begin{figure}[ht]
	\centering
	\includegraphics[width=\textwidth]{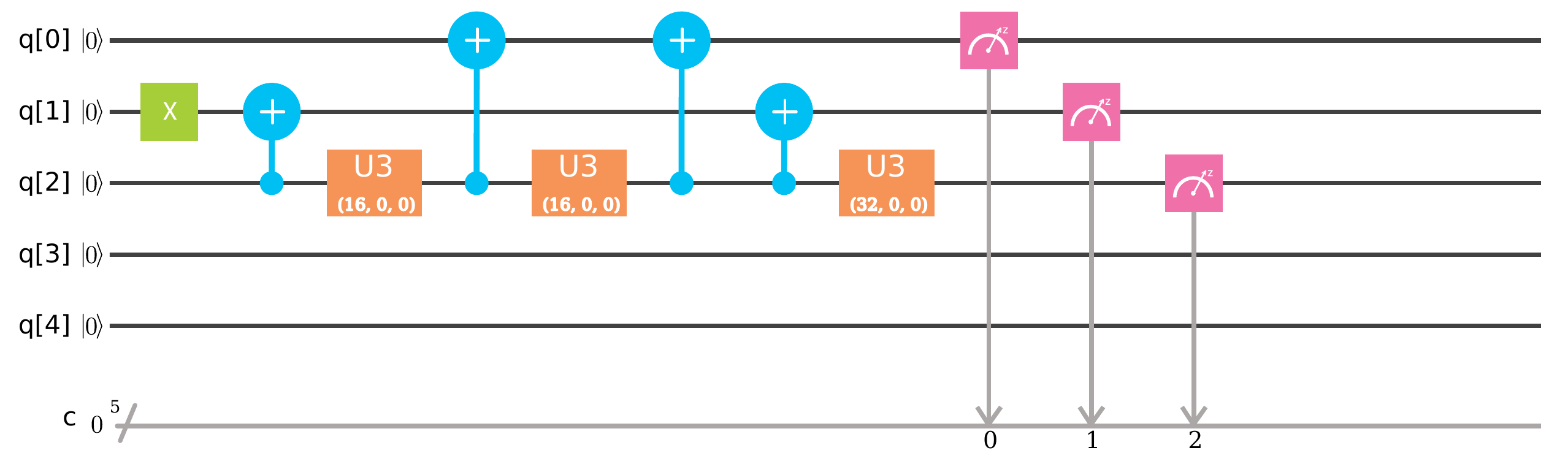}
	\caption{The quantum circuit for a large number of steps in the scaling limit (eq.\ \eqref{cSinglephasetd}) of a quantum walk in topological phase I for $N=4$ implemented on IBM-Q. In the circuit, $\textsf{U3(t,0,0)} = e^{-it Y}$.}\label{fig:NTContIBM}
\end{figure}

\begin{figure}[ht]
	\centering
	\includegraphics{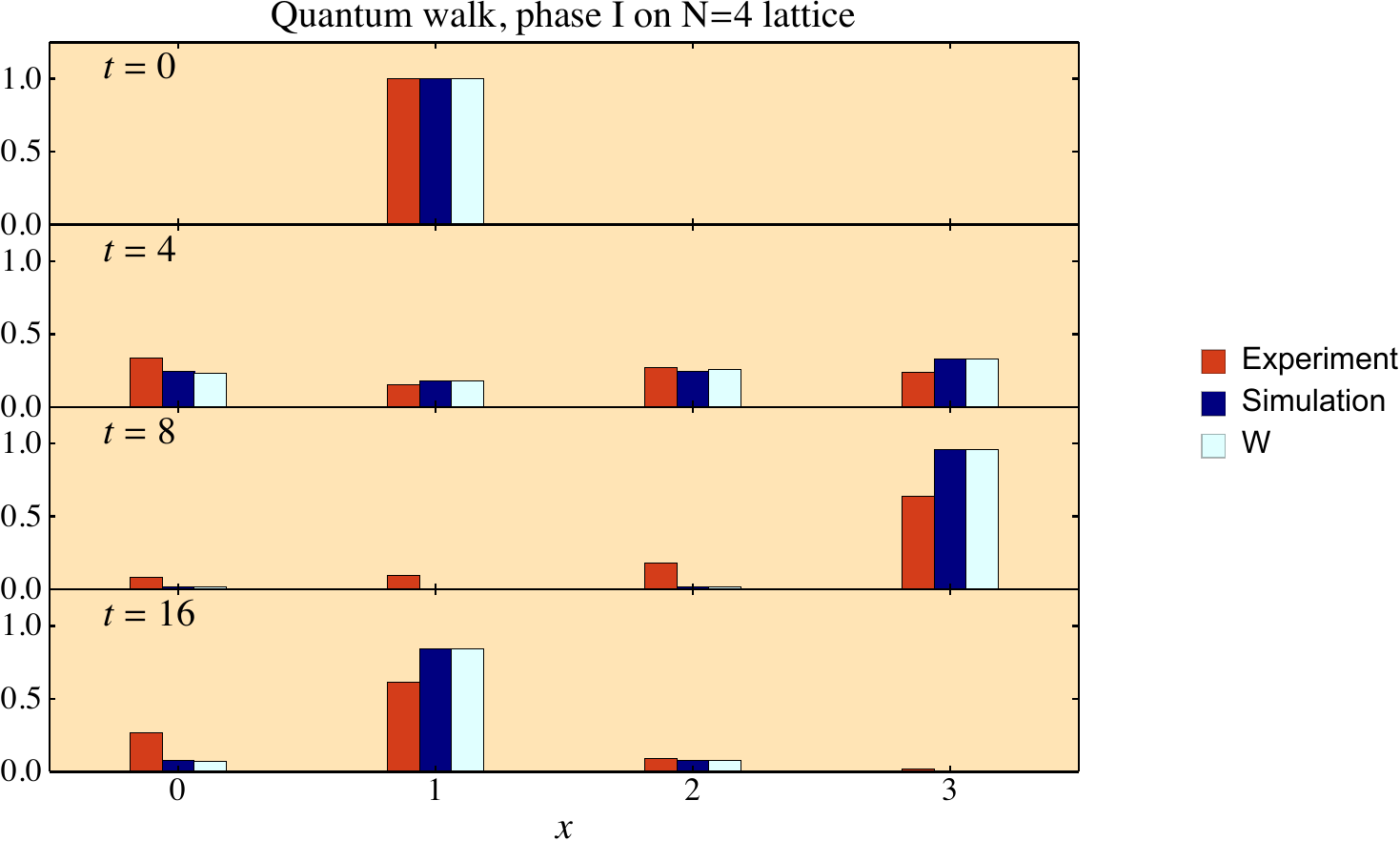}
	\caption{Quantum walk in topological phase I on a lattice of $N=4$ points with coin parameter $\epsilon = \frac{1}{32}$. Experimental results from IBM-Q demonstrate the absence of a bound state. They compare well with simulation, and repeated application of the matrix \eqref{eqW} representing a single step of the walker for number of steps $s$ related to $t$ via \eqref{eq17}.}\label{fig:NonTop}
\end{figure}

In this section we present experimental results from the IBM-QX4 five-qubit quantum computer. We implemented experimentally circuits for quantum walks on a lattice with $N=4$ points. The size of the lattice was limited by the number of available qubits, the quantum computer architecture (limitations on two-qubit quantum gate applications), and the total run time (limited by coherence time). Our algorithms used two qubits for the walker, and an additional qubit for the coin.

The quantum circuit for a single step $W$ (eq.\ \eqref{eqcW}) of a topological quantum walk in phases I/II with coin parameter $\epsilon = \frac{1}{8}$ implemented on IBM-Q is shown in figure \ref{fig:OneStepIBM}. The initial state is chosen so that it remains bound. Experimental results are compared with results from simulation in figure \ref{fig:OneStep}. The discrepancy is mainly due to a 4\% error in each CNOT quantum gate. Thus, the topological effect of a bound state is captured at a single step of the walker.

Performing more than a single step requires time which is longer than coherence time, and the algorithm fails on IBM-Q after the first step. To bypass this limitation, we go over to the scaling limit discussed in Section \ref{sec:3}.

The circuit in the scaling limit of a large number of steps of a topological quantum walk in phases I/II \eqref{eqcC} implemented on IBM-Q is shown in figure \ref{fig:ContIBM}. In figure \ref{fig:TimeDep}, experimental results are compared with results from simulation as well as by repeated application of a single step of the walker  $W$, with the number of steps related to the time parameter $t$ via \eqref{eq17}. For the choice of coin parameter $\epsilon = \frac{1}{32}$, we obtain good agreement between experiment and simulations. These results demonstrate the existence of a bound state at $x=1$ due to topological effects. The slight broadening of the peak at $x=1$ in the experiment is attributed to a 4\% error in the implementation of CNOT gates. 

For comparison, we simulated the evolution of a quantum walk in phase I where no topological effects are expected. We implemented the quantum circuit \eqref{cSinglephasetd} for a large number of steps and coin parameter $\epsilon = \frac{1}{32}$ on IBM-Q, as shown in figure \ref{fig:NTContIBM}. 
Starting from the same state with support at $x=1$, we observe a broadening of the peak after a finite time $t$, demonstrating the absence of a bound state, as expected. Experimental results are shown in figure \ref{fig:NonTop}. They compare well with results from simulation as well as repeated application of a single step of the walker  $W$. A small discrepancy is, once again, due to a 4\% error in CNOT gates.

It is worth noting that despite the limited resources of IBM-Q, our algorithm allowed us to implement a large number of walker steps and clearly demonstrate the existence of a bound state due to topological effects.

\section{Conclusion}
\label{sec:5}
We developed quantum circuits to efficiently evolve topological quantum walks on a finite lattice. Stroboscopic simulations, implementing an effective Hamiltonian, cast as quantum walks are a powerful tool in quantum simulations. Here, we implemented a continuous-time limit of a quantum walk to overcome the resource requirements demanded by stroboscopic simulations. We realized a logarithmic scaling for the lattice, polynomial growth in quantum gates, and an invariant relationship between the number of quantum gates, in a certain scaling limit,  and the number of steps of the walk. Our experimental implementation of the algorithm on the IBM-Q five-qubit quantum computer identified topological features in close agreement with our numerical simulations.

\acknowledgments{D.\ C.\ and G.\ S.\ thank the Army Research Laboratory, where part of this work was performed, for its hospitality and financial support.}

\end{document}